\documentstyle[11pt,paspconf,epsf]{article}

\def\kms{km~s$^{-1}$}
\def\cm3{cm$^{-3}$}

\begin{document}

\title{Ionized Gas in the Sgr~A Complex - VLA Observations of
       H168$\alpha$ and H270$\alpha$ Recombination Lines}
\author{K.R. Anantharamaiah$^{1,2}$, A. Pedlar$^3$ and W.M. Goss$^1$}
\affil{$^1$National Radio Astronomy Observatory, Socorro, NM 87801, USA}
\affil{$^2$Raman Research Institute, Bangalore 560 080, India}
\affil{$^3$Nuffield Radio Astronomy Laboratories, Jodrell Bank, UK}

\index{Source!Sgr~A}
\index{Galactic Center!HII}

\begin{abstract}
Radio recombination lines at $\lambda = 20$ cm reveals the presence of
an extended component of ionized gas of lower density ($\sim$ 100
\cm3) in the Sgr~A complex. This component extends well beyond the
thermal `mini-spiral' SgrA-West which seem to be embedded in it. The
low-density component is present over the entire extent of SgrA East
and possibly beyond and it is responsible for the turnover in the
spectrum of SgrA East and the halo observed by Pedlar et al (1989) at
$\lambda$ = 90 cm. The radial velocity of the extended ionized gas
range from +50 \kms\- to --200 \kms\- with minima of emission near --50 and
--150 \kms. The total mass of the low-density component is estimated to
be $\sim 10^4$ M$_\odot$. A possible extended feature at a high
velocity of 470 \kms\- is also detected, but needs further observations
to confirm it.
\end{abstract}

\section{Introduction}

At centimeter wavelengths, most of the radio continuum emission from
the Sgr~A complex arises from four distinct components: (1) the
ultra-compact non-thermal radio source SgrA* which may be at the very
center of the Galaxy (Lo et al 1998)), (2) the `mini-spiral' SgrA West
which is a rotating ionized ring around SgrA* (Ekers et al 1983,
Roberts and Goss 1993) (3) the shell-type, non-thermal source SgrA
East which is possibly an energetic supernova remnant, located
off-centered on SgrA* (Ekers et al 1983) and (4) a triangular shaped
halo which seem to surround all the other three components (Pedlar et
al 1989). In addition to these four main components, the SgrA complex
also contains many weaker but distinct thermal and non-thermal radio
features which have been classified and cataloged by Yusef-Zadeh and
Morris (1987)

Using high resolution observations at 90 cm and 20 cm, Pedlar et al
(1989) obtained constraints on the relative location along the line of
sight of the main components of the Sgr~A complex. They showed clearly
that much of the thermal gas associated with SgrA West lies in front
of SgrA* and SgrA East. Fig 1a shows a 20 cm image of the SgrA
region. The large shell like structure is SgrA East. The bright
compact source within the shell and the bright extended region around
it consists of SgrA* and SgrA West. At this resolution ($10''\times
8''$), SgrA* is not separated from SgrA West and the mini-spiral-like
structure of SgrA West is not resolved. Fig 1b shows an image of the
same region at 90 cm with the same angular resolution.  The difference
between the structures in Figs 1a and 1b is remarkable.  The change in
the structure is caused by the different emission mechanisms for SgrA
West and SgrA East (thermal and non-thermal) and their relative
location along the line of sight. The thermal source SgrA West is seen
as a depression in Fig 1b whereas it appears as a bright emission
feature superposed over the non-thermal source SgrA East in Fig
1a. The conclusion is obvious - SgrA West is in front of SgrA East and
the former has become optically thick at $\lambda$ = 90 cm and thus
absorbs the radiation from the non-thermal source SgrA East.

\begin{figure}[h]
\plotfiddle{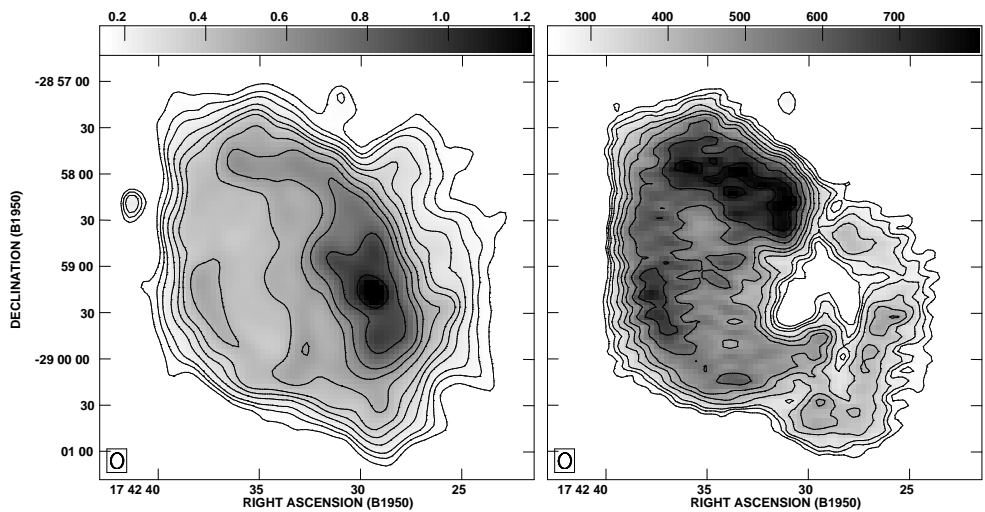}{2.5 in}{0}{75}{75}{-240}{-100}
\caption{(a) A 20 cm image of the SgrA region. Contour levels are
150, 175, 200, 250, 300, 350, 450 ... 1250 mJy/beam. (b) A 90 cm image
of the same region. Contour levels are 250, 275, 300, 350, 400, 450,
550, 650, 750 mJy/beam. The images are taken from Anantharamaiah et al
(1991) and convolved to a resolution of $10''\times 8''$.}
\end{figure}

The difference between the emissions at 20 cm and 90 cm is seen
quantitatively in Fig 2a, which displays EW cross-cuts through the
images in Figs 1a and 1b at the position of SgrA*. At 20 cm (thick
line) the minor peak on the east is due to the shell of SgrA East and
the strong peak is due to the combination of SgrA* and SgrA West. The
strong peak at 20 cm turns to an apparent absorption-like feature at
90 cm (thin line) indicating that both SgrA* and SgrA East are
absorbed by the optically thick SgrA West. An important implication of
this result is that, {\em it is not possible to obtain any measurement
of the flux density of SgrA* at frequencies below about 500 MHz} and
thus no constraints can be placed at these frequencies on theoretical
models which try to explain the overall spectrum of SgrA*
(e.g. Mahadevan 1998). At these frequencies, SgrA* is simply hidden behind
an absorbing screen.

\begin{figure}[h]
\plotfiddle{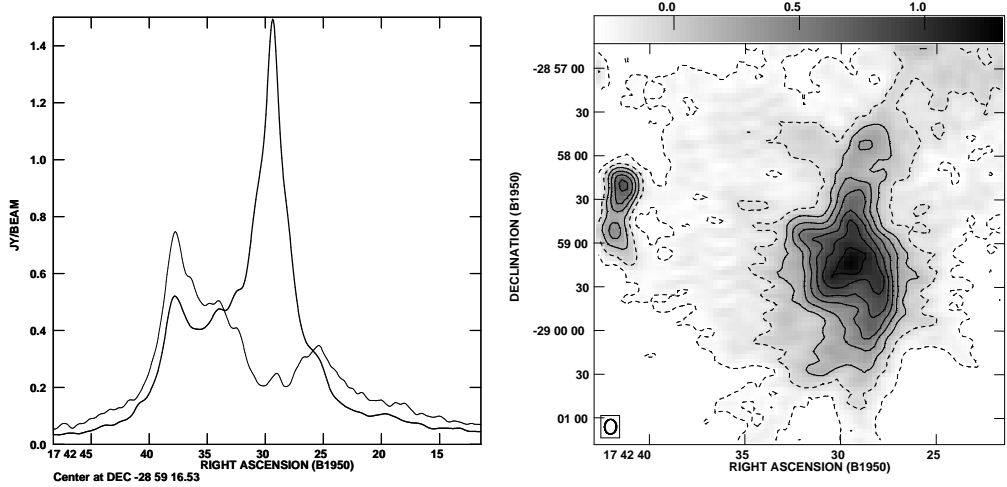}{2.5 in}{0}{75}{75}{-225}{-100}
\caption{(a) EW cross-cuts at 20 cm (thick line) and 90 cm (thin line) made 
from Fig 1 at the position of SgrA* (b) An image of the spectral index
$\alpha$ (where $S_\nu\propto\nu^\alpha$) of the SgrA region between
90 cm and 20 cm made from the images in Fig 1. The contour levels are
-0.3,-0.1,.1,0.3 ... 1.3.}
\end{figure}

Fig 2b shows an image of the spectral index $\alpha$ (where $S_\nu
\propto \nu^\alpha$) of the SgrA region between 20 cm and 90 cm made
from the images shown in Fig 1. As expected for optically thick
thermal gas, positive spectral indices are observed over the SgrA West
region. In fact the optically thick region seem to extend well beyond
the extent of SgrA West as determined by higher frequency
observations (Ekers et al 1983).  As noted by Pedlar et al (1989), the
mini-spiral SgrA West seems to be embedded in a somewhat lower density
extended ionized gas. Furthermore, the spectral index of SgrA East and
the halo between 20 cm and 90 cm is `flatter' ($\alpha \sim -0.3$ to
0.0) compared to the spectral index between 20 cm and 6 cm ($\alpha
\sim -1.1$ to $-0.9$: Ekers et al 1983, Pedlar et al 1989) indicating
that free-free absorption by thermal gas causes the turnover in their
non-thermal spectrum. A quantitative estimate of the amount of thermal
gas in front of SgrA East and also in the halo was made by Pedlar et
al (1989) using high resolution images at 2 cm, 6 cm, 20 cm and 90
cm. They showed that (1) the mini-spiral of SgrA West is embedded in a
thermal-halo, about 1.5 arcmin in angular extent and a ionized gas
mass of $\sim 500 M_{\odot}$ and (2) ionized gas with an emission
measures $\sim 10^5$ pc cm$^{-6}$ and $\tau_{90cm}\sim 1$ is present
over the extent of SgrA East ($\sim 3'$) and possibly beyond
and (3) the triangular-shaped halo surrounding both SgrA East and SgrA
West is possibly a mixture of thermal and non-thermal gas and may be
responsible for part or all of the spectral turnover of SgrA East
between 20 cm and 90 cm.

In this paper, we report observations of recombination lines at 20 cm
and 90 cm using the VLA. The 20 cm data clearly shows the presence of
non-zero velocity ionized gas present over the entire angular extent
of SgrA East and also beyond. The 90 cm data shows zero-velocity
lower-density ionized gas present along the line of sight to the Galactic
center.

\section{Observations of Recombination lines near 1374 and 332 MHz}

Recombination lines (RL) at lower frequencies (e.g. $\nu \le 1.4$ GHz)
are sensitive to lower density extended ionized gas due to enhanced
stimulated emission in such gas at higher quantum levels (Shaver 1975,
Pedlar et al 1978, Anantharamaiah 1985). On the other hand, at higher
frequencies ($\nu \ge 5$ GHz), it is easier to observe higher density,
compact ionized regions. High resolution recombination line
observations of the Galactic center region has so far been made only at
frequencies above 5 GHz.  (Van Gorkom et al 1985, Roberts and Goss
1993, Yusef-Zadeh et al 1995) and they have been useful in studying
relatively high density ($n_e > 10^3$ \cm3) ionized gas in SgrA West,
in the Arched filaments and in other compact HII regions in that
area. However, the extended  lower density ionized gas in the
SgrA complex, for which evidence was presented in the previous
section, has not been detected in high frequency recombination
lines. We therefore undertook observations of the H168$\alpha$ line at
$\nu_{rest}$ = 1374.6 MHz and the H270$\alpha$ line at $\nu_{rest}$ =
332.25 MHz using the VLA.

Observations at 1374 MHz were made in 1996 in both C and D configurations
of the VLA using a bandwidth of 6.25 MHz and 64 channels. The velocity
resolution was 21.3 \kms\- and the velocity coverage was $\pm$650
\kms. The data were edited, calibrated and imaged using standard
procedures in AIPS software. The angular resolution of the final line
and continuum images is $56.7'' \times 24.6''$ and PA = 35\deg.  For
the H270$\alpha$ line at 332 MHz, we used the data obtained in the D
configuration of the VLA in 1987 as a part of the continuum study
published by Pedlar et al (1989). The velocity resolution of the data
is 11.0 \kms\- and covers a range of $\pm$300 \kms. The angular
resolution is $8.2' \times 3.7'$ with PA = 10\deg.

\begin{figure}[h]
\plotfiddle{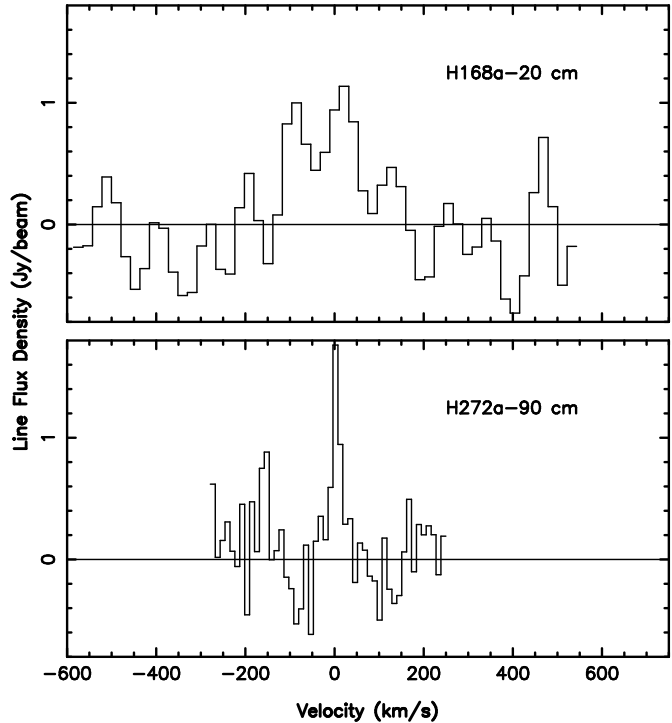}{3.5 in}{-90}{50}{50}{-200}{290}
\caption{H168$\alpha$ ({\em top}) and H270$\alpha$ 
({\em bottom}) recombination line profiles integrated over the SgrA 
complex. The narrow feature near -150 \kms\- in the lower profile
is the C272$\alpha$ line.}
\end{figure}

Fig 3 shows the spatially integrated profiles of the H168$\alpha$ and
the H270$\alpha$ lines. The two line profiles are different which
illustrate that recombination lines at different frequencies are
sensitivity to different components of ionized gas. The higher
frequency line (i.e. H168$\alpha$) extends over a large velocity range
($\sim$ 200 \kms) and arises in relatively higher density gas ($n_e >
100$ \cm3) close to the galactic center whereas the lower frequency
line (i.e. H270$\alpha$) is narrow ($\sim$30 \kms) and arises in a
lower density gas ($n_e < 10$ \cm3) which may be anywhere along the
line of sight (Anantharamaiah and Bhattacharya 1985). Note that the
expected radial velocity for the line-of-sight gas towards the
Galactic Center is 0 \kms. The narrow feature near $-$150 \kms\- in the
90 cm profile is the C27
0$\alpha$ line. The H270$\alpha$ and the
C270$\alpha$ lines are dominated by stimulated emission (Pedlar et al
1978, Anish Roshi and Anantharamaiah 1997) and thus their spatial
distribution follows that of the continuum. With the sensitivity of
the present observations, the lines are clearly detected only near the
continuum peak. The lower frequency line is not discussed further here
since it does not arise close to the Galactic center.

The H168$\alpha$ line profile in Fig 3 is consistent with single-dish
measurements at this position by Kestevan and Pedlar (1977). A slight
uncertainty in the baseline (i.e. the zero-level) is still visible
in the spectrum which is due to the residual 3-MHz ripple in the VLA
waveguide system. A significant narrow feature at a velocity of 470 \kms\-
is present in the H168$\alpha$ spectrum which needs to be confirmed by
further observations.

\begin{figure}[h]
\plotfiddle{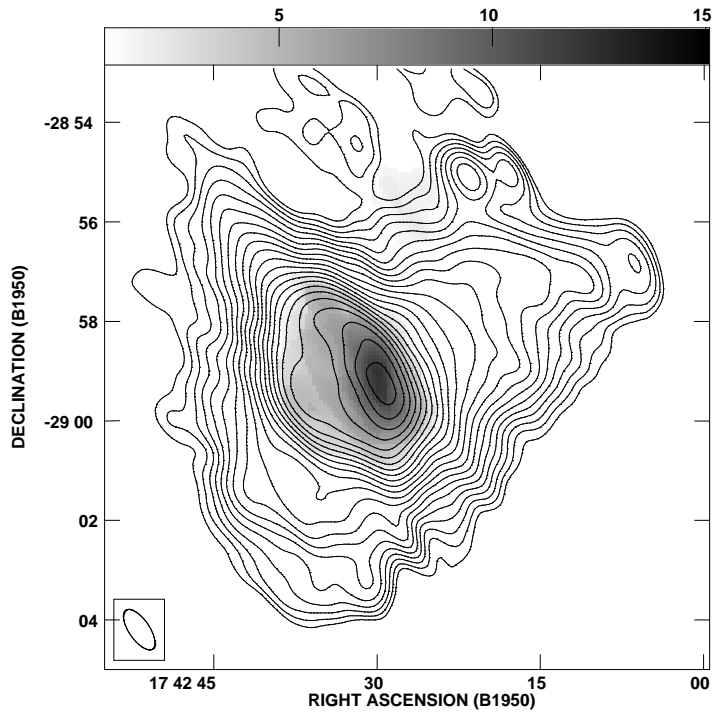}{3.75in}{0}{50}{50}{-160}{-70}
\caption{Continuum image of SgrA complex at 1.37 GHz in contours with 
H168$\alpha$ line emission integrated over the range +50 to --200 \kms\-
in grey scale. The first contour level is 0.3 mJy/beam and the
subsequent ones are successively higher by a factor of 1.2. The grey
scale range is 1 to 15 Jy/beam \kms.  The beam is $56.7''\times
24.6''$ and PA = 35\deg.}
\end{figure}

Fig 4 shows the continuum image of the SgrA complex at 20 cm (1374
MHz) in contours with H168$\alpha$ line emission integrated over the
velocity range $-$200 \kms\- to +42 \kms\- (moment 0) superposed in grey
scale. The angular resolution is $56.7'' \times 24.6''$ with PA = 35\deg.
The continuum image shows the three main components of the SgrA
complex: the triangular shaped outer halo which is about 10' in size,
the shell-like structure of SgrA East and the bright central region,
SgrA West. At this angular resolution, SgrA* is
not separated from SgrA West. The integrated line emission in grey
scale shows that the H168$\alpha$ line emission extends over the whole
of SgrA East. There can be little doubt that this extended emission
arises in ionized gas associated with the galactic center since the
velocities are non-zero and the velocity extent of this gas is much larger
than that of the line-of-sight gas detected in the H270$\alpha$
line. The detection of an extended region of H168$\alpha$ line
emission is consistent with the conclusions of Pedlar et al (1989) who
attributed the turnover in the spectrum of SgrA East to free-free
absorption by thermal gas.  There is also some weak H168$\alpha$ line
emission to the north of SgrA East where the base of the arched
filaments meet the SgrA halo at $(l,b)$ = (0,0).

\begin{figure}[h]
\plotfiddle{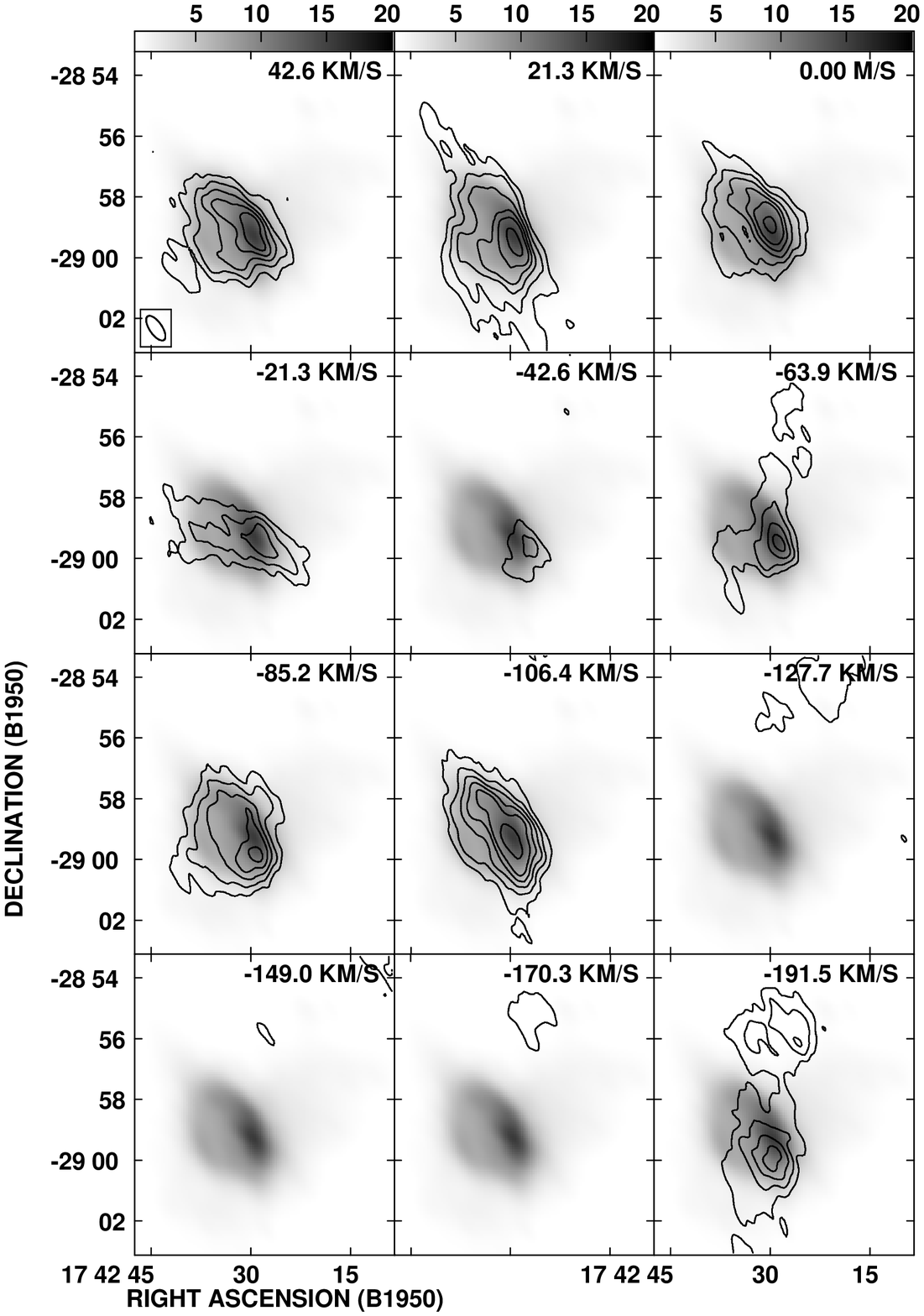}{4.75in}{0}{50}{50}{-170}{-30}
\caption{Channel maps of H168$\alpha$ emission in contours superposed
over a grey scale continuum image at 20 cm. The LSR velocities are
indicated in each frame. Contour levels are 20,30,40,50,60,80
mJy/beam. The beam size is as in Fig 3.}
\end{figure}

The velocities and spatial distribution of the H168$\alpha$ emission
can be seen in the channel maps shown in Fig 5. The H168$\alpha$
emission can be identified in Fig 5 in three distinct velocity ranges.
Taking into account the channel width of 21.3 \kms, these three
velocity ranges are (1) +50 to --10 \kms, (2) --75 to --115 \kms\- and
(3) --180 to --200 \kms. In addition to these emissions, a narrow high
velocity feature with $V_{lsr}=$ 470\kms\- was detected in the spectrum
shown in Fig 3. The channel image at $V_{lsr}=$ 470\kms\- is shown in
Fig 6 in contours superposed over the continuum image in grey
scale. If this line emission is confirmed by further observations,
then it represents in-falling ionized gas at a high velocity and
extended over a few parsecs.

In the velocity ranges +50 to -10 \kms\- and --75 to --115 \kms, the
line emission is spread over the entire extent of SgrA East with the
peak of emission in front of SgrA West.  Comparing the observed
velocities of the H168$\alpha$ line at the position of SgrA West with
the velocities observed at higher frequencies by Roberts and Goss
(1993) and Yusef-Zadeh et al (1995), it appears that the H168$\alpha$
line emission does not arise in the ionized gas that is directly
associated with SgrA West. The ionized gas detected in the
H168$\alpha$ line is clearly extended and possibly lies in front of
both SgrA East and West sources.  The similarity of the spatial
structure of line emission (which is from thermal gas) and that
of SgrA East (which is a non-thermal source) strongly suggests that
the H168$\alpha$ line is dominated by stimulated emission. It is
therefore possible that the line emission extends even beyond SgrA East
which may not be detected here due to lack of sensitivity and
diminished background continuum. Due to limited angular and velocity
resolution of these observations, no systematic motion of this gas can
be discerned. Since there are three distinct velocity ranges where
line emission occurs, the three regions are likely to be separated
along the line of sight. Some of this ionized gas may be associated
with the halo and mixed with the non-thermal gas as suggested by
Pedlar et al (1989).

\begin{figure}[h]
\plotfiddle{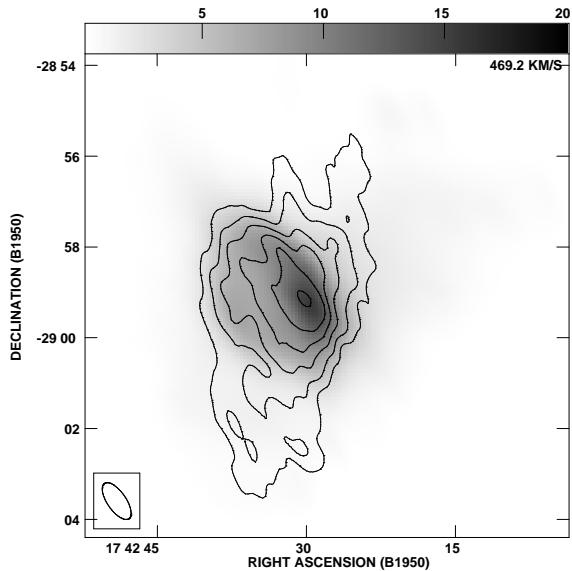}{2.5in}{0}{40}{40}{-130}{-55}
\caption{Channel image of the  H168$\alpha$ emission at V = 470 \kms\- 
(in contours) superposed over the continuum image at 20 cm. The contour
levels are 20, 30, 40, 50, 60, 80 mJy/beam. The beam size is  as
in Fig 4.}
\end{figure}

The peak of emission near --190 \kms\- occurs south of the continuum
peak. Some weak emission at this velocity also occurs slightly north of
SgrA East. Although negative velocities are observed in the higher
frequency H92$\alpha$ lines (Roberts and Goss 1993) in the southern
arm of SgrA West, the H168$\alpha$ line emission at this velocity is
more extended than the southern arm and thus may represent a distinct
component.

\section{Constraints on the Density and Mass of Ionized gas}

The extended ionized gas observed here in the H168$\alpha$ line is not
observed in the lower frequency H270$\alpha$ line nor in the higher
frequency H110$\alpha$ and H92$\alpha$ lines. High resolution
H110$\alpha$ and H92$\alpha$ VLA observations may also have missed
detecting  extended gas due to lack of sensitivity to such
structures. As explained in Section 1, lower frequency continuum
observations indicate the gas has a free-free optical depth $\tau_{ff}
\sim 1$ at $\lambda$ = 90 cm. These facts allow us to place some
constraints on the density and emission measure of the ionized gas.

\begin{figure}[h]
\plotfiddle{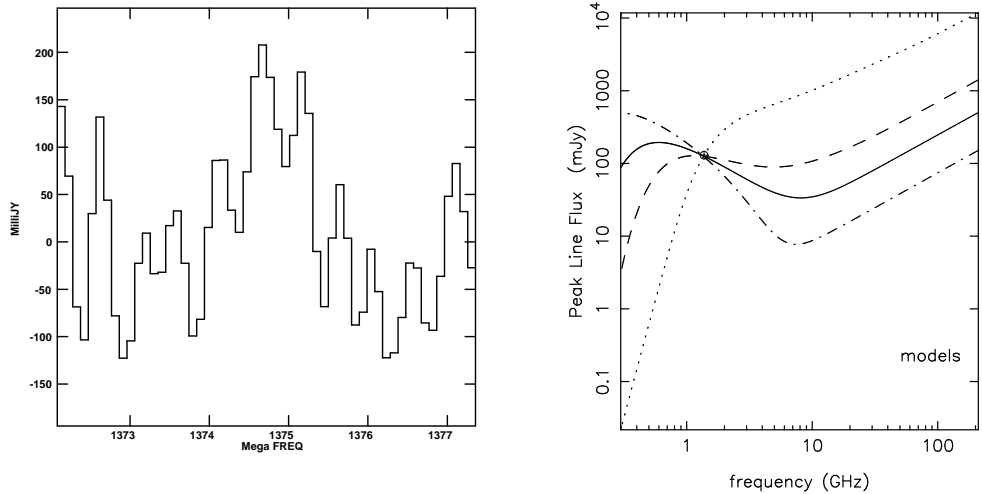}{3.0in}{0}{80}{80}{-220}{-100}
\caption{(a) H168$\alpha$ line profile integrated over a 
$54'' \times 135''$ region on SgrA East at the position RA =
$17^h42^m35.2^s$ and DEC = $-28^\circ 59'18.0''$. At this position
there is no contribution to the line from SgrA West. (b) Possible
models to account for the line emission. See text for parameters of the
models}
\end{figure}

Fig 7a shows a H168$\alpha$ line profile spatially integrated over a
$54'' \times 135''$ region of SgrA East centered at RA =
$17^h42^m35.2^s$ and DEC = $-28^\circ 59'18.0''$. This region is to
the east of SgrA West and thus the line profile has no contribution
from the latter. As discussed in Section 1, this region shows a
turnover in the spectrum of SgrA East which is attributed to free-free
absorption by thermal gas. It is reasonable to suppose that the
H168$\alpha$ line in Fig 7a arises in this gas. Although the line profile
in Fig 7a is complex consisting of at least 3 components, for the purpose
of illustration of the model, we
 approximate it by a Gaussian with a peak of 130 mJy/beam,
FWHM = 220 \kms\- and a centroid near 0 \kms. The integrated line flux
is $1.3 \times 10^{-21}$ W m$^{-2}$. The continuum flux density of
SgrA East over this region is $\sim$ 36 Jy/beam at 1.37 GHz. We
assume an intrinsic spectral index of $-0.9$ based on higher frequency
measurements (Ekers et al 1983). 

To illustrate possible models which can account for the observed line
emission, we assume a homogeneous slab of ionized gas at a temperature
of $10^4$ K to be in front of SgrA East. Fig 7b shows four models with
different densities.  The densities are 10 \cm3 for the dash-dot-dash
line, 100 \cm3 for the solid line, 500 \cm3 for the dashed line and
5000\cm3 for the dotted line. The emission measure in each model is
adjusted to fit the observed H168$\alpha$ line. It is clear from Fig
7b that the models with the highest (5000 \cm3) and lowest (10\cm3)
densities do not produce the observed behavior. The high density model
predicts a sharp increase in line strength towards higher frequencies,
whereas the lower density model predicts an increase towards lower
frequencies.  Only the intermediate density models predict lower line
strengths at both higher and lower frequencies as observed. We favor
the model represented by the solid line which has $n_e$ = 100
\cm3, EM = $3.3\times 10^5$ pc cm$^{-6}$. This model predicts a
free-free optical depth of $\sim$1.1 at 90 cm consistent with the
observed turnover in the spectrum of SgrA East (Pedlar et al
1989). Stimulated emission accounts for 90\% of the line strength. The
predicted line strengths in Fig 1 (solid line) are consistent with
non-detection at 330 MHz and between 5 GHz and 10 GHz. The model also
predicts that recombination lines at higher frequencies ($\nu>$ 10
GHz) may be detectable. It is possible that some of the H66$\alpha$
line emission near 22 GHz observed by Mezger and Wink (1986) from an
'extended' component is from this gas. In the model represented by the
solid line in Fig 1b, the total mass of the ionized gas is about $10^3$
M$_\odot$. Ionization of this gas can be maintained by a single O5
star. Note that this model, which is for illustration, is only for the
ionized gas which is in front of a part of SgrA East and which is
responsible for the line emission shown in Fig 7a. To account for the
total line emission shown in Figs 3 and 5, it is necessary to separate
the possible contribution from SgrA West. This separation is not
attempted here. If we assume that ionized gas with the same properties
extends over whole of SgrA East (i.e. about 4 arminutes), then the
total ionized mass is $\sim 8\times 10^4$ M$_\odot$.

The National Radio Astronomy Observatory is a facility of the National
Science Foundation operated under cooperative agreement by Associated
Universities, Inc.

\end{document}